\documentclass[sigconf]{acmart}

\usepackage{booktabs} 
\usepackage{algpseudocode}
\usepackage{amsmath}
\usepackage{color}
\usepackage{doi}
\usepackage{enumitem}
\usepackage{etoolbox}
\usepackage{fancyvrb}
\usepackage{float}
\usepackage{hyperref}
\usepackage{multicol}
\usepackage{pbox}
\usepackage{romannum}
\usepackage{tabularx}

\begin{document}
\title{Public Git Archive: a Big Code dataset for all}

\author{Vadim Markovtsev}
\orcid{0002-4235-0944}
\affiliation{%
  \institution{source\{d\}}
  \streetaddress{Calle de Claudio Coello, 16}
  \city{Madrid}
  \country{Spain}
  \postcode{28001}
}
\email{vadim@sourced.tech}

\author{Waren Long}
\orcid{0003-3310-6939}
\affiliation{%
  \institution{source\{d\}}
  \streetaddress{Calle de Claudio Coello, 16}
  \city{Madrid}
  \country{Spain}
  \postcode{28001}
}
\email{waren@sourced.tech}

\renewcommand{\shortauthors}{V. Markovtsev and W. Long}

\begin{abstract}
The number of open source software projects has been growing exponentially. The major online software repository host, GitHub, has accumulated tens of millions of publicly available Git version-controlled repositories. Although the research potential enabled by the available open source code is clearly substantial, no significant large-scale open source code datasets exist. In this paper, we present the Public Git Archive -- dataset of 182,014 top-bookmarked Git repositories from GitHub. We describe the novel data retrieval pipeline to reproduce it. We also elaborate on the strategy for performing dataset updates and legal issues. The Public Git Archive occupies 3.0 TB on disk and is an order of magnitude larger than the current source code datasets. The dataset is made available through HTTP and provides the source code of the projects, the related metadata, and development history. The data retrieval pipeline employs an optimized worker queue model and an optimized archive format to efficiently store forked Git repositories, reducing the amount of data to download and persist. Public Git Archive aims to open a myriad of new opportunities for ``Big Code`` research.
\end{abstract}

\copyrightyear{2018}
\acmYear{2018}
\setcopyright{rightsretained}
\acmConference[MSR '18]{MSR '18: 15th International Conference on Mining Software Repositories }{May 28--29, 2018}{Gothenburg, Sweden}
\acmBooktitle{MSR '18: MSR '18: 15th International Conference on Mining Software Repositories , May 28--29, 2018, Gothenburg, Sweden}
\acmDOI{10.1145/3196398.3196464}
\acmISBN{978-1-4503-5716-6/18/05}

\ccsdesc[300]{Human-centered computing~Empirical studies in collaborative and social computing}
\ccsdesc[300]{Software and its engineering~Software configuration management and version control systems}
\ccsdesc[100]{Software and its engineering~Software libraries and repositories}

\keywords{source code, git, GitHub, software repositories, development history, open dataset}

\maketitle

\section{Introduction}

Big code is revolutionizing software development. The revolution has begun with GitHub, whose collection of Git repositories is not just big, but vast: more than 24 million developers collaborating on over 67 million projects in 2017 \cite{octoverse}. GitHub has made version control accessible, therefore universal. The next stage of the revolution is permitting the automatic analysis of source code at scale, to support data-driven language design, to infer best (and worst) practices, and to provide the raw data to data hungry machine learning techniques that will be the basis of the next generation of development tools \cite{Sutton13, Barr16}. It requires source code archives that are both big and programmatically accessible for analysis.

The GHTorrent project \cite{Gousi13} took first steps in this direction, focusing on metadata in order to be scalable. Current source code datasets typically contain tens of thousands of projects at most \cite{Sutton13} and are dedicated to particular programming languages such as Java and JavaScript \cite{Raychev15}, thus lacking diversity and attracting critics \cite{Cosentino16}. Software Heritage \cite{dicosmo:hal-01590958} is a recent attempt to archive all the open source code ever written, however no public dataset has been published yet by them.

We present the Public Git Archive, the first big code dataset amenable to programmatic analysis at scale. It is by far the biggest curated archive of top-rated\footnote{We use ``top-rated``, ``top-starred``, `top-bookmarked` and ``having most stargazers`` interchangeably. The number of stargazers is a proxy on the degree of public awareness and project quality within the community.} repositories on GitHub, see Table \ref{dataset_stats} for comparison. The Public Git Archive targets large-scale quantitative research in the areas of source code analysis (SCA) and machine learning on source code (MLoSC). The dataset is made available via HTTP as a separate index file together with files in the Siva format, a novel archive format tailored for storing Git repositories efficiently \cite{siva}. Every GitHub repository can be forked; forks typically introduce subtle changes not necessarily merged into the origin. The naive way to obtain  forks is to clone them separately, requiring additional time and storage space. We describe the data retrieval pipeline which places forks into the original repository without mixing identities by reusing the existing Git objects \cite{6976092}. The dataset size becomes thus smaller, requiring users to download and store less data.

The main contributions of the paper are:
\begin{itemize}
\item The Public Git Archive dataset, which is the largest collection of Git repositories to date available for download.
\item The data retrieval pipeline which produces this dataset. Each part of that pipeline can scale horizontally to process millions of repositories.
\item The Siva repository archival format used in this dataset. This format allows to efficiently store forks.
\end{itemize}

\newcolumntype{M}[1]{>{\raggedright}m{#1}}

\bgroup
\def\arraystretch{1}
\begin{table*}
\begin{center}
\begin{tabular}{M{2.8cm} r r r r}
 & \textbf{Qualitas Corpus} \cite{QualitasCorpus:APSEC:2010} & \textbf{Sourcerer} \cite{6976088} & \textbf{GitHub Java Corpus} \cite{Sutton13} & \textbf{Public Git Archive}\tabularnewline
\hline
Number of projects  & 111 & 19,233 & 14,807 & 182,014\tabularnewline
Year of release & 2013 & 2014 & 2013 & 2018 \tabularnewline
Code language  & Java & Java & Java & 455 distinct \tabularnewline
Development history & No & No & No & Yes \tabularnewline
Number of files, $10^6$	  & 0.177 & 1.9 & 1.5 & 54.5 (HEAD) \tabularnewline
Lines of code, $10^6$  & 37.1 & 320 & 352 & 15,941 (HEAD) \tabularnewline
Storage size 	 & 1.3 GB & 19 GB & 14 GB & 3.0 TB \tabularnewline
\hline
\end{tabular}
\end{center}
\caption{Datasets comparison}
\label{dataset_stats}
\vspace*{-2.5em}
\end{table*}
\egroup

\section{Dataset production}

The dataset production consists of three steps, as follows.

\subsection{Compiling the list of repositories}

Similarly to existing research on GitHub mining \cite{Padhye:2014:SEC:2597073.2597113}, the focus of the dataset is put on the top-starred repositories. To compose the list of repository URLs, we make use of the metadata provided by GHTorrent \cite{Gousi13}: a scalable, queryable, offline database generated from listening events through GitHub API, and available to the research community as a data-on-demand service \cite{Gousi14}. The list for the Public Git Archive is based on GHTorrent's MySQL dump dated from January $1^{st}$, 2018.

We created a command-line application which streams the compressed GHTorrent MySQL dump, reads and processes the needed files and stores the intermediate tables on disk. This tool can also be used to filter repositories based on the number of stargazers and chosen programming languages by taking the intermediate tables for input. For the Public Git Archive, there is no filtering by language performed and the minimum number of stargazers is set to 50. The resulting list contains 187,352 unique Git repository URLs.

\subsection{Cloning Git repositories}

Once the list repository URLs is produced, we fetch the actual data with \texttt{borges} \cite{borges}: a container-friendly distributed system that clones and stores Git repositories at scale.

\begin{figure}[b]
\vspace*{-3.0em}
\begin{center}
\includegraphics[width=0.48\textwidth]{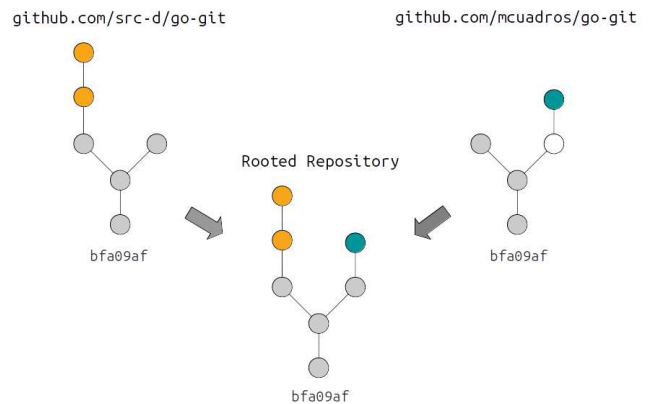}
\end{center}
\vspace*{-1em}
\caption{Rooted Repository}
\label{rooted_repo}
\end{figure}

Borges is designed as two separate standalone services: the producer reads URLs and determines which repositories should be processed next and adds new jobs for the consumer into the message queue; the consumer dispatches jobs to its thread worker pool. Multiple producers and consumers can be running; the message queue is also scalable. A job is a request to update a repository, new or existing. Each Git remote is fetched and each reference is pushed to the corresponding \textit{rooted repository}. This way, we store all references (including all pull requests) from different repositories that share the same initial commit -- \textit{root}, Fig. \ref{rooted_repo}. As a consequence, forks go into the same physical Git repository, thus being stored more efficiently.

Subsequently, Borges consumer's workers put Git packfiles belonging to rooted repositories into Siva files. Siva is an archival format \cite{siva} similar to \texttt{tar} and \texttt{zip}: it allows constant-time random file access, concatenation of the archive files, and seekable access to the contained files which are written verbatim—since packfiles are already compressed with \texttt{zlib}. Siva makes possible to store rooted repositories in an efficient and convenient way with minimal storage overhead. Internally, placing Git repositories inside Siva is implemented as \texttt{git push}. Borges can store Siva files in the local file system or Hadoop Distributed File System (HDFS).

The repositories belonging to the Public Git Archive were cloned in late January-February, 2018. A total of 8 consumers ran 32 threads each, taking one week under a 1Gbps internet connection. The "smart HTTP" Git protocol was used. The bulk download of 3.0 TB of data at the same connection speed takes under 8 hours — less than 1\% of the initial retrieval time normalized to single consumer. The exact storage space saved due to fork embedding is computationally expensive to calculate because a part of the pull requests are merged and the corresponding forks are not needed to be fetched. This is left for the future work.

From the initial 187,352 URLs, 3,156 had become inaccessible by the Git clone time, including 82 removed for legal reasons (HTTP 451 error messages returned by the server). The final amount of repositories cloned is 182,014 since several outliers \cite{git-bomb} could not be processed by our pipeline. 90\% of the repositories were cloned within the first 24 hours and they constitute 50\% of the final dataset size.

\subsection{Generating the index file}

Users of the dataset may prefer to work with a smaller subset of the terabytes collected of Siva files. A frequently observed use case is to triage files of certain programming languages. To address this and more preferences or restrictions, we generate from the finalized Siva files a CSV-type file including: per-repository metadata, detected license information, plus aggregate statistics on the number of files, lines and bytes per programming language. Each line in the CSV links to the corresponding Siva files, which in turn contain Git references of the corresponding repository. It becomes therefore possible to query the index file and choose which Siva files to download. The columns of the CSV file are explained in Table \ref{csv_columns}.

\section{Using Public Git Archive}

Links to the dataset download, as well as to all the relevant tools and scripts to reproduce it, are hosted in the official repository on GitHub\footnote{\href{https://github.com/src-d/datasets}{github.com/src-d/datasets}}. The Public Git Archive consists of (a) 248,043 Siva files (3.0 TB) with Git repositories and (b) the index file in CSV format. We also provide a command-line application to automate, accelerate and simplify downloading (a) and (b). In the columns of the CSV file, \textit{HEAD} is used to denote the latest commit of a branch; \textit{default HEAD} means the latest commit of the default branch. The default branch corresponds to the reference which is marked main on GitHub. Languages were detected using \textbf{enry} \cite{enry}. Licenses were detected using \textbf{go-license-detector} \cite{go-license-detector}. Lines of code were calculated using \textbf{gocloc} \cite{gocloc}. \textbf{GitHub API was not used}, as it is planned to extend the dataset to sources beyond GitHub.

\begin{figure}
\begin{center}
\includegraphics[width=0.48\textwidth]{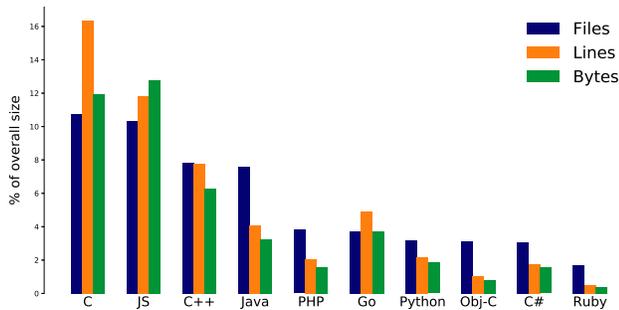}
\end{center}
\vspace*{-1.5em}
\caption{Statistics of 10 most popular programming languages in PGA.}
\label{langhist}
\end{figure}

Fig. \ref{langhist} shows the aggregated programming language statistics.

After the selected Siva files are downloaded, users can work with the dataset using \textbf{engine} \cite{engine}. The engine is an extension for Apache Spark which adapts Siva files as a Spark data source, allowing users to execute conventional Spark or PySpark queries to analyze Git repositories. It is also possible to unpack Git repositories from Siva files by using its Go language API or through the command line interface.

\section{Significance}

Analysis of source code has recently received significant research attention. The areas which can benefit from the Public Git Archive are statistical machine learning and natural language processing on source code. For example, source code modeling studies \cite{Sutton13, Barr16} have shown that the performance of $n$-gram models critically depends on the dataset size. That's why the presented dataset can enhance research in topics like automatic naming suggestion \cite{Barr15}, program prediction \cite{Raychev15}, topic modeling and semantic clustering \cite{Kuhn07}, bug detection \cite{gao:icse:2017}, and automated software transpilation \cite{barr:issta:15}. It can also provide valuable insights for compiler developers into how the programming languages are used by the open source community.

Another promising research direction is inter-project source code clone detection. Social programming platforms with minimal boundaries between projects like GitHub have facilitated code reuse across multiple projects. A number of studies has been carried out about those ecosystems in the recent years \cite{Nguyen:2013:SRC:3107656.3107682}. Code clones were first studied within single projects, but as GitHub grew further, different reasons for the appearance of duplicated code snippets have been explored, e.g. ``accidental`` clones due to imprecise API usage protocols  \cite{1541846} or automatic program repair \cite{6035728}. The Public Git Archive enables the research community to study source code clones across project boundaries, not limited to a single language and having large graphs with over 10,000 projects \cite{Gharehyazie:2017:HCC:3104188.3104225}.

\bgroup
\def\arraystretch{1.3}
\begin{table}[t]
\vspace*{-1.0em}
\begin{center}
\begin{tabular}{M{2.9cm} M{5.4cm}}
\textbf{Column name} & \textbf{Description} \tabularnewline
\vspace*{-1.0em}
\rule{8.6cm}{.4pt}
$url$ & URL of the GitHub repository. \tabularnewline
$siva\_filenames$ & Siva files which contain parts of that repo. \tabularnewline
$file\_count$ & Number of files in default HEAD reference. \tabularnewline
$langs$ & Languages encountered in default HEAD. \tabularnewline
$langs\_$$\{byte,lines,files\}$\\$\_count$ & Byte, line, file counts per each language, in the same order as \textit{langs}.\tabularnewline
$commits\_count$ & Number of unique commits in the Siva files which refer to that repository. \tabularnewline
$branches\_count$ & Number of references, tags excluded. \tabularnewline
$fork\_count$  & Number of remotes in the referring Siva files. \tabularnewline
$\{empty,code,comment\}$\\$\_lines\_count$ & Number of empty, code, commented lines in default HEAD. \tabularnewline
$license$ & License names and corresponding confidences. \tabularnewline
\vspace*{-1.5em}
\rule{8.6cm}{.4pt}
\end{tabular}
\end{center}
\vspace*{-1em}
\caption{CSV columns}
\label{csv_columns}
\vspace*{-3em}
\end{table}
\egroup

\section{Updates}

In order to evolve along the constantly changing open-source landscape, The Public Git Archive needs to be regularly updated. Several technical challenges arise from this requirement. The typical way to organize dataset updates is to provide regular snapshots, as GHTorrent does. However, every snapshot of our dataset would require considerable disk space. The solution is to manage incremental updates consisting of the differences from the previous snapshot. Two ways to implement this solution are:

\begin{itemize}
\item To pull changes into every packfile in every Siva file of the dataset. The \texttt{git pull} operation requires the whole new packfile to be read, and this is precisely what one would like to avoid.
\item To generate binary diffs of the Siva files. However, diffing Git packfiles is not straightforward. They are compressed and even a single Git object which is removed at the beginning of a packfile changes the whole binary stream, making it necessary to retain the old objects which are no longer referenced in the new packfile. GitHub always returns a single packfile during \texttt{git clone} and runs garbage collection from time to time, effectively breaking the binary diffs.
\end{itemize}

The challenges of the first implementation are harder to resolve technically. The second implementation seems more feasible and has ongoing research. The current plan to update the Public Git Archive is to publish complete snapshots, limiting their lifetime. There are going to be Long Term Support (LTS) snapshots with extended lifetime and researchers are encouraged to focus on them. The exact schedule is subject to change and is updated on the Public Git Archive website.

\section{Privacy and licensing}

The Public Git Archive contains the full commit history for each public repository, including commit messages, timestamps, author names and emails. GitHub Terms of Service (GHTS) explicitly allow passing such public information to third parties as long as the goal is doing research or archiving \cite{ghtos}. The Public Git Archive is maintained solely for research purposes, and the collected credentials are not to be used in any way except as allowed by the GHTS.

Despite the public nature of the information, some developers may prefer to take their projects down or private, making repositories inaccessible as noticed in section \Romannum{2}. We do provide a communication channel for repository removal requests, the full details provided on the official Public Git Archive website.

Each rooted repository inside Siva files is licensed separately and according to the manifested project license. Projects which do not have an explicit license are distributed under the same terms as stated in GHTS exclusively for research purposes. The index file is licensed under a \textit{Creative Commons Attribution-NonCommercial-ShareAlike 4.0 International} license.

\section{Limitations}

Dataset miners should take into account several potential threats to validity \cite{KGBSGD16}.

Regarding the data collection process and the traditional trade-off between freshness and curation \cite{Cosentino16}, we choose to emphasize the curation of the dataset rather than a limited amount of data up-to-date. We rely on GHTorrent for the list of repositories to retrieve, thus our update schedule depends on the upstream. Consequently, the dataset and the pipeline to collect it are entirely transparent. The output is never exactly the same, though, as GitHub is a dynamic environment and repositories may change or become inaccessible over time.

Other notable concern is about the generalization of the dataset. Selecting repositories based on the number of stargazers is arguable and may introduce bias. Fair probabilistic sampling of the complete list of repositories should improve the diversity, e.g. stratified random sampling \cite{Nagappan:2013:DSE:2491411.2491415}. Other popularity indicators can be explored, as the number of forks or accepted pull requests \cite{Sutton13}. By focusing on the number of stars as a measure of people's interest and awareness of the project, there is a risk to miss quality samples. As a result, various source code quality metrics should be considered. Finally, there will be duplicate files across different repositories \cite{Lopes:2017:DMC:3152284.3133908}. Mentioned suggestions constitute the basis for the future work.

\section{Conclusion}

In this paper, we presented the Public Git Archive, the largest source code dataset of top-starred Git repositories, described the novel scalable pipeline to reproduce it and the tooling to download and use it. The Public Git Archive is made available through HTTP and includes the source code of the projects, their metadata, and their development history. The retrieval pipeline is efficient and the dataset size is optimal thanks to the distributed cloning system and the custom Git repository archive format. We believe that the Public Git Archive — over ten times larger than any of the currently available datasets — has the potential to boost the quality, confidence and diversity of the software engineering and mining research.

\bibliographystyle{ACM-Reference-Format}
\bibliography{bibliography}

\end{document}